\documentclass[a4paper, 11pt]{article}

\usepackage[latin1]{inputenc}
\usepackage{graphicx,color}
\usepackage{courier}
\usepackage{amsmath,amsfonts}
\usepackage{amssymb,dcolumn,exscale}
\usepackage[small]{caption}
\usepackage[width=16.6cm, left=2.2cm,top=2.5cm,height=23.7cm]{geometry}
\usepackage{cite}
\usepackage{psfrag,booktabs,longtable,subfig}

\newcommand{\as}{\alpha_s}
\newcommand{\w}{\omega}
\newcommand{\gE}{\gamma_E}

\newcommand{\p}{\partial}
\newcommand{\mq}{m_{\widetilde{q}}}
\newcommand{\mg}{m_{\widetilde{g}}}
\newcommand{\shat}{\hat{s}}
\newcommand{\qqbar}{\ensuremath{q \bar{q}}}
\newcommand{\gluinopair}{\ensuremath{\widetilde{g}\,\widetilde{g}}}

\newcommand{\GeV}{\ensuremath{\,\mathrm{GeV}}}
\newcommand{\TeV}{\ensuremath{\,\mathrm{TeV}}}

\newcommand{\pb}{\ensuremath{\,\mathrm{pb}}}

\newcommand{\CI}{C_{\bf I}}
\newcommand{\DI}{D_{\bf I}}

\newcommand{\lnNt}{\ln(\widetilde N)}

\newcommand{\non}{\nonumber}
\newcommand{\ord}{{\mathcal O}}

\def\shat{{\hat s}}
\def\muf{{\mu^{}_f}}
\def\mufs{{\mu^{2}_f}}
\def\mur{{\mu^{}_r}}
\def\murs{{\mu^{2}_r}}
\def\muh{{\mu^{}_h}}
\def\mus{{\mu^{}_s}}

\allowdisplaybreaks[3]

\begin{document}
\begin{titlepage}
\noindent
DESY 13-035 \hfill May 2013\\
LPN 13-021 \\
SFB/CPP-13-16\\
\vspace{2.3cm}

\noindent 
\begin{center}
\Large{\bf
Phenomenology of QCD threshold resummation\\[1ex] for gluino pair production at NNLL
}\\
\vspace{1.5cm}
\large
Torsten Pfoh\\[10mm]
\normalsize
{\it 
Deutsches Elektronensynchrotron DESY\\[1mm]
Platanenallee 6, D--15738 Zeuthen, Germany \\[2mm]
\texttt{\footnotesize{torsten.pfoh@desy.de}}
}
\vspace{2.3cm}

\large
{\bf Abstract}
\vspace{-0.2cm}
\end{center}
We examine the impact of threshold resummation for the 
inclusive hadronic production cross section of gluino pairs at
next-to-next-to-leading-logarithmic accuracy, compared to the exact
next-to-leading-order cross section and the next-to-next-to-leading-order
approximation.
Here, we apply formulas derived recently in the classical Mellin-space formalism.
Moreover, we give the analytic input for the alternative momentum-space formalism
and discuss the crucial points of the numeric implementation.
We find that soft resummation keeps the hadronic cross section close to the
fixed next-to-leading-order result. 
\vfill
\end{titlepage}

\newpage
\section{Introduction}
Within the search for new physics at the LHC experiments,
one hopes to find evidences for heavy color-charged particles.
These appear in many scenarios for physics beyond the
Standard Model (SM). A prominent candidate is the gluino
which is a Majorana fermion and the superpartner of the SM gluon 
within various realizations of Supersymmetry (SUSY), as one of the 
most favorite models.
Unfortunately, the direct search at ATLAS and CMS have only produced 
exclusion limits\footnote{See Refs.~\cite{:2012rz,:2012gq} for example.} 
so far which, of course, depend on the model under
consideration, and the assumptions on the SUSY particle spectrum.

In order to separate possible signals from the SM background,
a crucial theoretical input quantity is the inclusive hadronic production
cross section $\sigma_{pp\to \gluinopair X}$. As known from standard 
Quantum Chromodynamics (QCD), the latter is given by a sum over the 
various partonic production channels where the partonic cross sections 
$\hat{\sigma}_{ij \to \gluinopair}$ are convoluted with their respective 
parton luminosity functions $L_{ij}$. Explicitly, one has 
\begin{eqnarray}
  \label{eq:totalcrs}
  \sigma_{pp\to \gluinopair X}(s,\mg^2,\mq^2,\mufs,\murs) &=&
  \sum\limits_{i,j = q,{\bar{q}},g} \,\,\,
  \int_{4\mg^2/s}^1 \,
  d\tau \,\, L_{ij}(\tau, \mufs)\,\,
  \hat{\sigma}_{ij \to \gluinopair} (\tau s,\mg^2,\mq^2,\mufs,\murs)
  \, ,
\end{eqnarray} 
where the parton luminosities are themselves given by a
convolution of the parton distribution functions (PDFs) 
\begin{eqnarray}
  \label{eq:partonlumi}
  L_{ij}(\tau, \mufs) &=&
   \int_0^1 dx_1 \int_0^1 dx_2\,\delta(x_1 x_2 - \tau)
  f_{i/p}\left(\mufs,x_1\right)
  f_{j/p}\left(\mufs,x_2\right)
  \, ,
\end{eqnarray} 
and we have introduced the hadronic center-of-mass (cms)
energy $s$, the partonic cms energy $\shat=\tau s$, the
factorization scale $\muf$, and the renormalization
scale $\mur$. The gluino mass is denoted by $\mg$, and we
assume mass degeneracy among the squarks flavors, therefore using 
a single scale $\mq$.

At leading order (LO) in perturbation theory, gluino pair production 
is driven by the partonic sub processes of gluon fusion 
$g g  \to \widetilde{g} \widetilde{g}$ and quark-antiquark annihilation
$q\bar{q} \to \widetilde{g} \widetilde{g},\ q = d,u,s,c,b$.
At next-to-leading-order (NLO), also the $gq$ channel opens~\cite{Beenakker:1996ch}. 
Close to the threshold however, its contribution is suppressed compared to gluon fusion 
and quark-antiquark annihilation. The full NLO result has been implemented in the public program
 \texttt{Prospino}~\cite{Beenakker:1996ed}.

As very well known, 
the cross section develops powers of large logarithms $\ln(\beta)$ 
near the production threshold, where the velocity 
\begin{eqnarray}
\label{eqn:beta}
\beta &=& \sqrt{1-4\mg^2/\shat}\ \ \equiv\ \ \sqrt{1-\rho}
\end{eqnarray}
of the produced particle pair goes to zero. 
The so-called threshold logarithms spoil the validity of the
perturbative expansion in the strong coupling constant $\alpha_s$. 
However, they can be resummed systematically to all orders in 
perturbation theory, where the quantity $\alpha_s\ln(\beta)$ counts as order one.
In this context, threshold logarithms are sometimes called soft logarithms and 
one talks about soft resummation. 
There are two approaches to soft resummation which we will shortly discuss below. 
We will refer to them as Mellin-space formalism \cite{Sterman:1986aj,Catani:1989ne}
and momentum-space formalism \cite{Becher:2006nr,Becher:2006mr}.

Another difficulty arises through the exchange of soft gluons between the final state 
particles. At fixed order in perturbation theory, this gives rise to so-called
Coulomb terms proportional to powers of $\alpha_s/\beta$.
Obviously, with decreasing $\beta$, the above ratio becomes $\ord(1)$
and should therefore be resummed as well.
This can be done in the framework of non-relativistic QCD (NRQCD).
A joined soft and Coulomb resummation has been worked out in 
the momentum-space formalism \cite{Beneke:2010da}. In the Mellin-space formalism
however, one would need to calculate the Mellin transformations of the NRQCD expressions
analytically (or at least semi-analytically as a function of the Mellin moments),
a problem which has not been solved so far. Therefore, Coulomb terms are
included at fixed order only.

For any process, soft resummation up to so-called next-to-leading-logarithmic (NLL) 
accuracy requires the knowledge of the color-decomposed Born cross section
with respect to the $SU(3)_c\,$-color configuration of the produced
(s)particle pair. At next-to-next-to-leading-logarithmic (NNLL) precision,
one needs the color decomposition of the NLO cross section near the threshold,
where higher powers of $\beta$ are skipped.
For gluino pair production, the state of the art in the momentum-space
formalism is a combined soft and Coulomb resummation up to NLL accuracy~\cite{Falgari:2012hx}. 
Within the Mellin-space approach, the NLL results of Ref.~\cite{Kulesza:2009kq,Beenakker:2009ha}
have been extended to NNLL precision recently in Ref.~\cite{Langenfeld:2012ti}.
Bound-state and finite-width effects at fixed order have been discussed in 
Ref.~\cite{Hagiwara:2009hq,Kauth:2011vg}.
In the context of NLL resummation, finite-width effects have been discussed recently 
in Ref.~\cite{Falgari:2012sq}. It has been found that for $\Gamma/\mg\approx 5\%$, 
the effect is of the order of the ambiguities due to higher-order contributions in the 
resummation formula. In this article, we work in the 
zero-width approximation which is very well justified if the squark and neutralino masses 
are about the order of the gluino mass.
If one expands the NNLL threshold resummation formula up to $\as^2$ (modulo the
factor $\as^2$ in the Born cross section), one reproduces the threshold logarithms up to
next-to-next-to-leading-order (NNLO). One further obtains constant terms which
are generically different at NNLO for the two approaches. However, a proper matching
eliminates these constants and one is left with the threshold approximation 
($\text{NNLO}_{\rm th}$), where, after factorization of the Born cross section, all 
NNLO constants are set to zero.
At NLO, the constants are kept as these refer to the exact fixed order result.
 
If the gluino is heavy (of the order of $1\TeV$), the threshold-enhanced terms
give the dominant contribution at given order in perturbation theory.
Combining the full NLO result with the $\ord(\as^2)$ threshold enhanced 
terms to the NNLO approximation ($\text{NNLO}_{\rm approx}$), one finds an
increase of the cross section of about $20\%$ with respect to the fixed NLO 
result \cite{Langenfeld:2012ti}. 
The question of interest is how these findings change in the presence of NNLL resummation. 
Although the resummation formula for soft logarithms has been derived analytical, it has
not been implemented yet. In this context, a general question is how to deal with ambiguities 
related to the choice of scales or the treatment of the Landau pole in momentum integrals.
A different source of uncertainty comes from the unknown squark masses.
However, the hadronic process is mainly driven by gluon fusion 
which has only a weak dependence on $\mq$. As a consequence, the impact of the squark masses
on the cross section is negligible to good approximation. In the threshold limit,
the dependence on $\mq$ actually vanishes.
The impact of higher order contributions is estimated by a variation of the
factorization and renormalization scale.
As discussed in Ref.~\cite{Langenfeld:2012ti}, the main source of uncertainty comes from 
the shape of the gluon PDFs at high momentum fraction $x$, and the value of $\alpha_s(M_Z)$.
The possible large discrepancies are not covered by the individual PDF errors,
and give the main uncertainty when setting a reliable exclusion limit.

In this article, we study the impact of NNLL soft resummation in the Mellin-space approach.
In the sections~\ref{sec:Mellin} and~\ref{sec:beta}, we briefly review the methods 
for soft (and Coulomb) resummation in the Mellin- and momentum-space formalism.
Concerning the latter, we derive the color-decomposed hard function at NLO which is needed for 
NNLL resummation. 
We discuss the ingredients of the potential function which collects all results 
from NRQCD derived from the NLO non-relativistic potential. 
In the sections~\ref{sec:implN} and ~\ref{sec:implbeta}, we discuss technical aspects of the 
numerical implementation of the resummation formulas in the two approaches, and explicitly 
apply the Mellin-space formalism. The inclusive hadronic cross section is discussed in
section~\ref{sec:pheno} and we give our conclusion in section~\ref{sec:conc}. 
An explicit formula for the matching of the momentum-space resummation formula onto the 
NNLO approximation is provided in the appendix.

\section{Threshold resummation in Mellin space}
\label{sec:Mellin}
 
The traditional approach to threshold resummation has been invented 
in Ref.~\cite{Sterman:1986aj,Catani:1989ne}, see also Ref.~\cite{Catani:1990rp,Contopanagos:1996nh,Catani:1996dj,Catani:1996yz,Kidonakis:1997gm,Moch:2005ba}
for further discussions.
Resummation is performed in Mellin space after introducing
moments $N$ with respect to the variable $\rho = 4\mg^2/\shat\,$
of momentum space. Neglecting the dependencies on $\muf$ and $\mur$, one has
\begin{eqnarray}
  \label{eq:mellindef}
  \hat{\sigma}_{ij}^N(\mg^2) &=& {\rm\bf M}[\hat{\sigma}_{ij}(\mg^2)](N)\;=\,
  \int\limits_{0}^{1}\,d\rho\, \rho^{N-1}\,
  \hat{\sigma}_{ij}(\shat,\mg^2)\, ,
\end{eqnarray}
where the threshold limit $\beta\to 0$ corresponds to $N\to\infty$. The resummation of
threshold logarithms is achieved by the formula 
\begin{eqnarray}
  \label{eq:sigmaNres}
 \hat{\sigma}_{ij,\, {\bf I}}^{\rm res,\,N}      
  &=&
  \hat{\sigma}^{B,\,N}_{ij,\, {\bf I}}\,
  g^0_{ij,\, {\bf I}} \; g^{0\,C}_{ij,\, {\bf I}}(N+1) \,
  \exp \Big[ G_{ij,\,{\bf I}}(N+1) \Big] +
  {\cal O}(N^{-1}\ln^n N) \, .
\end{eqnarray}
All contributions from soft and collinear radiation exponentiate
and are collected in the function $G_{ij,\,{\bf I}}$. To
NNLL accuracy, it may conveniently be expanded according to
\begin{eqnarray}
  \label{eq:GNexp}
  G_{ij,\, {\bf I}}(N) =
  \lnNt \cdot g^1_{ij}(\widetilde\lambda)  +  g^2_{ij,\, {\bf I}}(\widetilde\lambda)  +
  a_s\, g^3_{ij,\, {\bf I}}(\widetilde\lambda)  + \dots\, ,
\end{eqnarray}
where $\widetilde N \equiv N\exp(\gamma_E)$ with $\gamma_E$ denoting the Euler-Mascheroni constant, 
$\widetilde\lambda \equiv a_s\,\beta_0\, \ln \widetilde N$, $\beta_0$ as defined in Ref.~\cite{Langenfeld:2012ti},
and $a_s \equiv {\alpha_s }/{(4 \pi)}$.
The explicit expressions are calculated by a double integral over a set of 
anomalous-dimension functions. Some of these depend on the $SU(3)_c\,$-color configuration 
of the final-state gluino pair which we label by a capital index ${\bf I}$. 
The color-summed partonic cross section for a given channel is simply given by
\begin{eqnarray}
\label{eq:colorsum}
\hat{\sigma}_{ij \to \gluinopair} &=& 
\sum\limits_{{\bf I}}\, \hat{\sigma}_{ij,\,{\bf I}} 
\, .
\end{eqnarray}
The possible final-state color configurations are obtained by a decomposition
of the initial color states into irreducible representations according to 
\begin{eqnarray}
  \label{eq:colgg}
 \mathbf{8}\times\mathbf{8} &=& \mathbf{1}_s + \mathbf{8}_s + \mathbf{8}_a +
  \mathbf{10} + \mathbf{\overline{10}} + \mathbf{27}_s
  \, ,\\
  \label{eq:colqq}
 \mathbf{3}\times\mathbf{{\bar 3}} &=& \mathbf{1}_s + \mathbf{8}_s + \mathbf{8}_a\,,
\end{eqnarray}
for gluon fusion and quark-antiquark annihilation, respectively. 
There are symmetric ($s$) and anti-symmetric ($a$)
color states. The corresponding quadratic Casimir operators $\CI$,
which show up in the color-decomposed partonic cross section, take the values
\begin{eqnarray}
\CI &=&\lbrace 0,\,3,\,6,\,8\rbrace \qquad\text{for}\qquad 
{\bf I}\,=\,\lbrace {\bf 1},\,{\bf 8},{\bf 10},\,{\bf 27} \rbrace
\end{eqnarray}
independent from the symmetry properties.
The (color-decomposed) Born cross section $\hat{\sigma}^{B}_{ij,\, {\bf I}}$ factors out
in Eq.~(\ref{eq:sigmaNres}) which is also true for the fixed order partonic cross section 
in the threshold limit $\hat \sigma_{ij,\, {\bf I}}^{\rm th}$.
The matching constant $g^0_{ij,\, {\bf I}}$ collects all contributions independent 
of $N$ and has to be determined order by order in perturbation theory by 
expanding\footnote{This produces all the threshold logarithms of the form $\ln^k(N)$.}  
the right-hand side of Eq.~(\ref{eq:sigmaNres}) in $\alpha_s$ and matching onto 
$\hat \sigma_{ij,\, {\bf I}}^{\rm th}$.
Coulomb corrections are not resummed by the above ansatz. They are formally
treated as part of the hard scattering function on which the exponent has to be
matched. Therefore, one introduces a second matching term
$\,g^{0\,C}_{ij,\, {\bf I}}$ \cite{Kawamura:2012cr} as a power series in $\alpha_s$ 
which is actually a function of $N$. If Coulomb corrections are not considered, 
$g^{0\,C}_{ij,\, {\bf I}}$ reduces to a factor one.

Note that starting from NNLO, one has non-relativistic corrections of non-Coulomb type 
which cause the cross section to depend on the spin-configuration $S$ of the produced 
heavy particle pair 
\cite{Czarnecki:1997vz,Beneke:1999qg,Czarnecki:2001gi,Pineda:2006ri,Beneke:2009ye}.
These terms are also not treated by the exponent~(\ref{eq:GNexp}) and thus are fully hosted
by the hard matching constant. As they also depend on the Mellin variable $N$, it is 
convenient to include them into $g^{0\,C}_{ij,\, {\bf I}}(N)$, although they do not arise
from Coulomb exchange. Alternatively, one could also introduce a third matching constant
$g^{0\,NC}_{ij,\, {\bf I}}(N)$. The spin dependence is described by a parameter 
$v_{\rm spin}$ which is zero for the gluino pair being in a spin singlet (S=0) 
configuration and $-2/3$ for a spin triplet (S=1), see Ref.~\cite{{Beneke:2009ye}}. 
At the threshold, gluon fusion gives rise to a singlet, where quark-antiquark annihilation produces a triplet.
All ingredients of the NNLL threshold resummation in Mellin space for gluino pair production
can be found in Ref.~\cite{Langenfeld:2012ti}.

\section{Threshold resummation in momentum space}
\label{sec:beta}

An alternative approach which features resummation in momentum space has been
invented in Ref.~\cite{Becher:2006nr,Becher:2006mr} in the context of deep
inelastic scattering. Here, soft and collinear radiation are described
in the framework of soft-collinear effective theory. 
In Ref.~\cite{Beneke:2010da}, this method has been adopted to heavy (s)particle 
pair production and extended to the summation of Coulomb corrections.
The starting point is the observation that near the threshold the partonic 
production cross section can be factorized into a hard function $H$, a soft function 
$W$ which collects soft fluctuations, and a potential function $J$ which sums Coulomb exchange. 
Below the threshold, one has a discrete spectrum of bound-states. If these are taken 
into account, $J$ should be expressed as a function of the energy $E=\sqrt\shat -2\mg$ 
relative to the threshold \cite{Beneke:2010da,Beneke:2011mq,Falgari:2012hx}.

At fixed order in $\as$, the partonic cross section close to the threshold
can be written as~\cite{Beneke:2010da}
\begin{eqnarray}
\label{eq:sigmafac}
\hat{\sigma}_{ij \to \gluinopair} ({\shat},\mu) &=&
\sum_S\sum_{\bf I} H_{ij,\,{\bf I}}^S(\mu)\,\int_0^\infty d\w\,
J_{\bf I}^S(E-\frac\w 2)\,W_{\bf I}(\w,\mu)\,,
\end{eqnarray}
where one sums over the spin and color configurations of the final-state
particle pair. The spin dependence enters the hard and the potential function at NNLO first. 
In order to account for the different scales of the problem and to
achieve resummation of threshold logarithms, one calculates the hard function at the hard scale 
$\mu_h$ and the soft function at soft scale $\mu_s$.
Then, one solves renormalization group equations for the hard and the soft function
which, in the latter case, sums the soft logarithms. 
The solutions are used to evolve $H(\mu_h)$ and $W(\mu_s)$ to 
the common scale $\muf$ which is used for the convolution of the partonic cross sections 
with the parton luminosities.
The  general all order resummation formula reads~\cite{Beneke:2010da} 
\begin{eqnarray}
\label{eq:sigmares2}
\hat{\sigma}_{ij \to \gluinopair}^{\rm res} ({\shat},\muf) &=&
\sum_S\sum_{\bf I} H_{ij,\,{\bf I}}^S(\muh)\,
U_{ij,\,{\bf I}}(\mg,\muh,\mus,\muf)\\
&&\times
\int_0^\infty d\w\,\frac{J_{\bf I}^S\bigl(E-\frac{\w}{2}\bigr)}{\w}
\,{\left(\frac{\w}{2\mg}\right)}^{2\eta}
\tilde s_{ij,\,{\bf I}}\left(2\ln\left(\frac\w\mus\right)+\p_\eta,\mus\right)
\frac{e^{-2\gamma_E\, \eta}}{\Gamma(2\eta)}\,,\non
\end{eqnarray}
where the summation of Sudakov-double logarithms is included in the evolution function 
$U_{ij,\,{\bf I}}$ which is given by
\begin{eqnarray}
\label{eq:U}
U_{ij,\,{\bf I}}(\mg,\muh,\mus,\muf) &=& 
\left(\frac{4\mg^2}{\muh^2}\right)^{-2a_\Gamma(\muh,\mus)}
{\left(\frac{\muh^2}{\mus^2}\right)}^\eta\times
\exp\biggl[4\bigl(S(\muh,\muf)-S(\mus,\muf)\bigr)\quad\\
&&\ -2a_{ij,\,{\bf I}}^V(\muh,\mus)
+ 2a_i^\phi(\mus,\muf) + 2a_j^\phi(\mus,\muf)\Biggr]\,.\non
\end{eqnarray}
The function $\tilde s_{ij,\,{\bf I}}(\rho,\mu_s)$ is the 
Laplace transform of the $\overline{\text{MS}}$-renormalized soft function 
$\overline{W}_{\bf I}(\w,\mu)$ which, to NLO accuracy, is given by 
\begin{eqnarray}
\label{eq:softfunc}
\tilde s_{ij,\,{\bf I}}(\rho,\mu) &=& 1 +
a_s(\mu)\*\left[(C_i+C_j)\*\left(\rho^2+\frac{\pi^2}{6}\right) 
- 2\,\*\CI\,\*(\rho-2)\right] + \ord\left(a_s^2\right)\,,
\end{eqnarray}
where we again define $a_s\equiv \as/(4\pi)$.
The auxiliary variable $\eta$ in the  Eq.~(\ref{eq:sigmares2})
is set to $\eta=2a_\Gamma(\mus,\muf)$ after performing the derivative.
It contains single logarithms \cite{Falgari:2012hx} which can be seen by
expanding $\eta= 4\, a_s (C_i+C_j)\ln(\mus/\muf)+\ord(a_s^2)$.
From the latter expression one also learns that $\eta$ is negative and
tends towards zero as $\mus$ approaches $\muf$. This behavior is 
preserved at higher orders in $\as$. As a consequence, the integration kernel
$(\w/(2\mg))^{2\eta}/\w$ has to be understood in a distributional sense 
\cite{Becher:2006mr} as will be further discussed in Sec.~\ref{sec:implbeta}.

The quadratic Casimir operators $C_i$ and $\CI$ depend on the color configuration
of the initial and final states, respectively.
For gluino pair production via gluon fusion we have $C_i=C_j=C_A=3$,
for quark-antiquark annihilation we have $C_i=C_j=C_F=4/3$. 
With $\mus\leq\muf$ it hence follows that $\eta\leq 0$.
The function $a_\Gamma(\nu,\mu)$ and the Sudakov exponent $S(\nu,\mu)$   
are given to NNLO precision in Eq.~(86) and Eq.~(87) of Ref.~\cite{Becher:2007ty}.
For NNLL resummation, the respective NLO expressions are sufficient.
For $i=j$, these are functions of expansion coefficients $\Gamma_k^i$ of the cusp 
anomalous dimension
\begin{eqnarray}
\label{eq:gammacusp}
\Gamma_{\rm cusp}^i &=& \sum_{k=0}^\infty \Gamma_k^i\, {a_s}^{k+1} 
\;\equiv\;C_i\,\gamma_{\rm cusp}\,,
\end{eqnarray}
and coefficients of the QCD beta-function $\beta_k$.
The latter are listed in Eq.~(85) of the above article, whereas the cusp anomalous
dimension is given for the quark-antiquark channel in Eq.~(81). For gluon fusion, one
just has to multiply these results by $C_A/C_F$ according to the Casimir scaling
indicated in Eq.~(\ref{eq:gammacusp}) above. 
For $i\neq j$, one would replace 
\begin{eqnarray}
\Gamma_{\rm cusp}^i &\rightarrow & \Gamma_{\rm cusp}^{ij}
= \frac 12\,(\Gamma_{\rm cusp}^i+\Gamma_{\rm cusp}^j)
\equiv\frac 12\,(C_i+C_j)\sum_{k=0}^\infty\gamma_{\rm cusp}^{(k)}\, {a_s}^{k+1} \,.
\end{eqnarray}
Note also that the coefficients $\Gamma_k^i$ coincide with the coefficients 
$A_i^{(l)}$ with $l=k+1$, used in the Mellin-space formalism 
(see Ref.~\cite{Langenfeld:2012ti,Moch:2008qy}).  
The functions $a_{ij,\,{\bf I}}^V$ and $a_{i}^\phi$ in Eq.~(\ref{eq:U}) are obtained
in analogy to $a_\Gamma$ by a replacement of $\Gamma_{\rm cusp}^{ij}$ with 
$\gamma_{ij,\,{\bf I}}^V=\gamma_i+\gamma_j+\gamma_{H,\,\bf I}$ in the first case,
and with $(\gamma_i^\phi+\gamma_j^\phi)/2$ in the second. The single particle anomalous dimension 
functions $\gamma_i$ are related to soft radiation from the (mass-less) initial-state
particles and are collected in appendix~A of Ref.~\cite{Becher:2009qa}. 
The anomalous dimension $\gamma_{H,\,\bf I}$ refers to soft radiation connected
to a massive final-state particle in the color representation ${\bf I}$.
NLO results are given in Eqs.~(3.31,32) of Ref.~\cite{Beneke:2009rj}.
Here, one can also find expansion coefficients of $\gamma_i^\phi$ which are related 
to the evolution of the parton distribution function.
Explicit expressions up to NLO for color-triplets and octets 
are given in Eqs.~(D.17-20).

When taking results from the literature, care has to be taken with respect to
the meaning of the parameter $n_f$. Typically, $n_f$ denotes
the number of active flavors which contribute to the evolution of $\alpha_s$.
So it does in the articles listed above where the various coefficients of
the QCD beta function and anomalous dimensions are collected.
However, in some papers about gluino pair production at the threshold\footnote{See 
Ref.~\cite{Kauth:2011vg,Langenfeld:2012ti} for instance.} $n_f$ denotes the total amount
of quark (and squark) flavors which, depending on the renormalization scheme,
may differ from the number of active flavors. 

As discussed in Ref.~\cite{Beneke:2010da}, the all-order resummation 
formula~(\ref{eq:sigmares2}) does not depend on the hard and soft scale 
$\muh$ and $\mus$. However, as all ingredients are only 
known to finite order in $\alpha_s$, a truncation of the perturbative expansion 
induces a residual dependence on these scales. A good perturbative behavior is obtained 
by a suitable choice for $\mu_h$ and $\mu_s$.
For the case of gluino pair production, these are $\mu_h\approx k_h\mg$ and 
$\mu_s= k_s \mg\beta^2$ with $k_h$ and $k_s$ being numbers of $\ord(1)$.
The choice of a running soft scale $\sim \mg\beta^2$ is also necessary to reproduce the
threshold logarithms at NNLO (fixed order) \cite{Beneke:2011mq}.
In practice, the above choice becomes problematic when $\beta$ 
tends zero. Moreover, $\as(\mus)$ approaches the Landau pole.   
Therefore, one should introduce a parameter $\beta_{\rm cut}$ \cite{Beneke:2011mq}. 
For $\beta>\beta_{\rm cut}$, a running soft scale is used such that the perturbative 
expansion is not spoiled by large logarithms. 
On the other hand, for $\beta<\beta_{\rm cut}$, the scale is fixed and kept
in the perturbative regime.  
Note that a special treatment of the Landau pole is also required 
in the Mellin-space approach when performing the Mellin inversion (see 
Refs.~\cite{Catani:1996dj,Catani:1996yz} for instance).
In the momentum-space approach a proper choice of $\beta_{\rm cut}$ should keep the 
ambiguities which arise when matching the resummed to the fixed order cross section as small
as possible. For gluino pair production, this has been analyzed in Ref.~\cite{Falgari:2012hx}.
Finally, we note that the choice of the soft scale actually determines what is being
resummed. A dedicated discussion can be found in Ref.~\cite{Bonvini:2013td}.
Within the Mellin-space formalism on the other hand, one has the implicit scale choices 
$\mus=\mg/N$ and $\muh=\muf$, and the two approaches are formally identical up to 
$\ord(1/N)$-terms (see for instance the discussion in Ref.~\cite{Becher:2007ty}). 

Coming back to the momentum-space approach, one requires the potential function
$J_{\bf I}^{S}(E)$ which describes the exchange of soft gluons between the final-state 
particles as well as higher order corrections coupled to these diagrams. 
At fixed order in perturbation theory, pure Coulomb corrections (which 
correspond to ladder diagrams) give rise to terms proportional to $(\as/\beta)^n$.
As already mentioned, they can be treated by methods of NRQCD using the LO 
non-relativistic potential and have been found to be summed by a Sommerfeld 
factor $\Delta^C$ \cite{Fadin:1990wx}. Starting from NLO, 
one also has to deal with terms multiplied by $\as(\as/\beta)^n$. Their summation requires 
the non-relativistic potential at NLO \cite{Beneke:1999qg,Beneke:2009ye}. The LO and NLO 
Greens functions of the non-relativistic Schr\"odinger equation, $G^{(0)}_{C,\,\bf I}(E)$
and $G^{(1)}_{C,\,\bf I}(E)$, are given in Ref.~\cite{Beneke:2010da,Beneke:2011mq},
and the contributions to the potential function are obtained by taking two times the 
imaginary parts.
For positive values of the energy $E$ relative to the threshold, one finds the simple 
leading expression
\begin{eqnarray}
\label{eq:JI0}
J_{\bf I}^{(0)}(E) &=&  
 \frac{\mg^2}{2\pi}\,\sqrt{\frac{E}{\mg}}\,
\Delta^C\left(\pi\as\sqrt{\frac{\mg}{E}}\,\DI\right)\theta(E)\\
& = &  \frac{\mg^2}{2\pi}\,\sqrt{\frac{E}{\mg}}\,
\left[1 - a_s\,2\pi^2\DI\sqrt{\frac{\mg}{E}} + a_s^2\,\frac {4\pi^4\DI^2}{3}\,\frac{\mg}{E}
+ \ord\left(a_s^3\right) \right]
\theta(E)\,,\non
\end{eqnarray}
where the Sommerfeld factor is given by $\Delta^C(x)= x/(\exp(x)-1)\,$,
and we have introduced the color coefficient $\DI$ of the QCD potential
\begin{eqnarray}
\DI = \frac 12\, C_{\bf I} - C_A \,=\,\lbrace -3,-3/2,\,0,\,1 \rbrace
\qquad 
\text{for}\quad {\bf I}\,=\,\lbrace {\bf 1},\,{\bf 8},{\bf 10},\,{\bf 27} \rbrace\,.
\end{eqnarray}
Note that close to the threshold where $E\ll 2\mg$, one has 
$\sqrt{\mg/E}=1/\beta + \ord(\beta)$ with $\beta\ll 1$.
This gives rise to the expression $\Delta^C\left(\pi\as\,\DI/\beta\right)$
also described in Ref.~\cite{Kulesza:2009kq,Langenfeld:2012ti}
which makes the summation of $1/\beta$-terms evident.
For $\DI<0$ the potential is attractive, for $\DI>0$ it is repulsive. 

The NLO Coulomb function $J_{\bf I}^{(1)}(E)$ can not be brought to a simple form.
However, it can be expressed as a linear combination of dilogarithms and
nested harmonic sums with a complex argument 
$\lambda(E) = \as(-\DI)/(2\sqrt{-E/ \mg})$ \cite{Beneke:2011mq}.
As a first approximation, it may be sufficient to include $J_{\bf I}^{(1)}$
at fixed order in $\as$. It reads
\begin{eqnarray}
\label{eq:J1fix}
J_{\bf I}^{(1)}(E)= \frac{\mg^2}{2\pi}\, a_s^2\, 2\pi^2\DI\biggl[
\beta_0\biggl(\ln\Bigl(\frac{E}{\mg}\Bigr) + 2\ln\Bigl(\frac{2\mg}{\mu_C}\Bigr)\biggr)
-a_1 \biggr] + \ord\left(a_s^3\right)
\end{eqnarray}
with $a_1=31/9\;C_A-10/9\; n_l\,$. 
The choice of the scale $\mu_C$ is discussed in detail in Ref.~\cite{Beneke:2011mq}.
The complete fixed order non-relativistic
corrections at NNLO are obtained by adding to Eq.~(\ref{eq:J1fix})
the convolution of the $\ord(\as)$-Sommerfeld term with the NLO threshold approximation
(without the Coulomb term), as well as a non-Coulomb contribution discussed below. 
The full result has been given in Ref.~\cite{Beneke:2009ye} first.

If the potential is attractive, discrete bound states may develop
below the threshold. Ignoring the gluino widths, these are described by
\begin{eqnarray}
J_{\bf I}^{\rm bound}(E) =\, 2\sum_{n=1}^\infty \delta(E-E_n)
{\left(\frac{\mg\as(-\DI)}{2n}\right)}^3
\left(1 + \frac{\as}{4\pi} \delta r_1\right)\theta(-\DI)\,,\qquad\qquad (E<0)\,,
\end{eqnarray}
for S-wave production \cite{Beneke:2010da,Beneke:2011mq}, where the bound-state
energies are given by
\begin{eqnarray}
E_n =\, -\frac{\mg\as^2\DI^2}{4n^2}\left(1 + \frac{\as}{4\pi} e_1\right)\,.
\end{eqnarray}
The correction terms $\delta r_1$ and $e_1$ stem from the NLO Coulomb
Greens function \cite{Beneke:2005hg} and can be found in Ref.~\cite{Beneke:2011mq}, 
Eqs.~(B.4) and (B.5). We stress that bound-state production
($2\to 1 $ process) has different kinematics opposed to pair production  
($2\to 2 $ process) and therefore requires different fixed order formulas
already at the Born level. 
The NLO cross section for gluinonium production has been studied in Ref.~\cite{Kauth:2011vg}. 

As mentioned above, the non-relativistic potential also exhibits spin-dependent
terms of non-Coulomb type starting from NNLO. 
Due to explicit results for the threshold enhanced part of the NNLO
cross section \cite{Langenfeld:2012ti}, we know
that these terms play a minor role  for gluino pair production. 
They comply with around $5\%$ of the total NNLO contribution to the partonic cross section. 
Thus, it is seems to be sufficient to include these terms at fixed order which has also been 
done for top-quark pair production \cite{Beneke:2011mq}. We follow the latter reference and define 
the factor
\begin{eqnarray}
\Delta_{\rm NC,\,{\bf I}}^S(E) &=& \DI
\bigl(C_A - 2\DI(1 + v_{\rm spin})\bigr)\,\theta(E)\,,
\end{eqnarray}
where the parameter $v_{\rm spin}$ has been defined above.
In summary, the potential function required for NNLL resummation is given by
\begin{eqnarray}
J_{\bf I}^S(E)=J_{\bf I}^{(0)}(E)\,\Bigl(1 + \as^2 \ln(\beta)
\Delta_{\rm NC,\,{\bf I}}^S(E)\Bigr) + J_{\bf I}^{(1)}(E)\,,
\end{eqnarray}
where $\ln(\beta) \approx \ln(\sqrt{E/\mg}) $.\\

As last ingredient of the NNLL resummation formula, the hard function 
$H_{ij,\,{\bf I}}^S(\muh)$ is needed at NLO.
At LO, it is related to the partonic Born cross section 
\cite{Beneke:2011mq,Beneke:2010da,Falgari:2012hx}
\begin{eqnarray}
\hat \sigma_{ij,\,{\bf I}}^{(0)}(\muh)
&\approx & \frac{\mg^2}{2\pi}\,\sqrt{\frac{E}{\mg}}\,H_{ij,\,{\bf I}}^{(0)}(\muh)
\;\approx\; \frac{\mg^2}{2\pi}\,\beta\,H_{ij,\,{\bf I}}^{(0)}(\muh)\,,
\end{eqnarray}
where the approximations are valid in the limit $\shat\to 4\mg^2$.
Near the threshold, the LO cross section is factored out from
higher order contributions, where, apart from an overall factor $\beta$ due to the 
Born term, all positive powers of $\beta$ are set to zero.
Writing
\begin{eqnarray}
\label{eq:norm0}
H_{ij,\,{\bf I}}(\muh) = H_{ij,\,{\bf I}}^{(0)}(\muh) 
\left[ 1 + a_s(\muh)\*H_{ij,\,{\bf I}}^{(1)}(\muh)
+ a_s^2(\muh)\*H_{ij,\,{\bf I}}^{(2)}(\muh) + \ord\left(a_s^3\right)\right],
\end{eqnarray}
the required NLO hard functions are obtained
by comparing the NLO fixed order threshold results with the NLO expansion
of Eq.~(\ref{eq:sigmares2}). 
For this purpose, it is sufficient to set $J_{\bf I}^S=J_{\bf I}^{(0)}$,
and we require the various anomalous dimension functions
at LO only. Using $\gamma_{\rm cusp}^{(0)}=4$, 
$\gamma_{H,\,\bf I}^{(0)}=-2 \CI$, $\gamma_i^{\phi(0)}=-\gamma_i^{(0)}$, 
and $E=\mg\beta^2$, we find in agreement with Eq.~(D.3) from Ref.~\cite{Beneke:2010da}
\begin{eqnarray}
\label{eq:sigmaresNLO}
\hat{\sigma}_{ij \to \gluinopair}^{\rm res,\,exp} ({\shat},\muf) &=&
\sum_{\bf I} \frac{\mg^2}{2\pi}\,\beta\,H_{ij,\,{\bf I}}(\muh)\;
\Biggl[
1 + a_s \Biggl(-\frac{2 \DI \pi^2}{\beta} + 
    4 (C_i + C_j) \ln^2(8 \beta^2) \\
&& \quad - 4\Bigl(\CI + (C_i + C_j) \bigl(4 + L_{f\tilde g}\bigr)\Bigr)\ln(8 \beta^2)
    + (C_i + C_j) L_{f\tilde g}^2  \non \\
&& \quad + 2\Bigl(\CI + (C_i + C_j) \bigl(4 - L_{fh}\bigr)\Bigr)L_{f\tilde g} 
   + 12\CI + (C_i + C_j)\left(32-\frac{11}{6}\pi^2\right)\non\\
&& \quad -\left(2\CI - 4(C_i + C_j)\ln(2) 
   -\bigl(\gamma_i^{(0)} + \gamma_j^{(0)}\bigr)\right) L_{fh}
    + (C_i + C_j) L_{fh}^2\Biggr)  + \ord\left(a_s^2\right) \Biggr],\non
\end{eqnarray}
where we have defined
$L_{fh} = \ln(\muf^2/\muh^2)$, and $L_{f\tilde g} = \ln(\muf^2/\mg^2)$.
The dependence on the soft scale has canceled between the soft function and the 
expansion of the evolution function. The terms proportional to $L_{fh}$
run the hard function $H_{ij,\,{\bf I}}^{(1)}(\muh)$ down to the factorization scale 
$\muf$, see also the discussion in Ref.~\cite{Beneke:2010da}. 

The explicit evaluation of the soft and the potential function at NLO
has produced all threshold enhanced terms proportional to $\ln^k(8\beta^2)$
($k=1,2)$ and $1/\beta$ which are easily identified within the fixed order
cross section in the threshold limit, given in the
Refs.~\cite{Beenakker:1996ch} and \cite{Langenfeld:2012ti}.
However, the renormalization scale has been set equal to the factorization scale
in the latter articles. Adopting this choice simplifies the right-hand side of 
Eq.~(\ref{eq:sigmaresNLO}) due to $L_{fh}=0$. 
From now on, we set $\muh=\muf=\mu$, $\as(\mu)\equiv\as$, and switch to the notation of 
Ref.~\cite{Langenfeld:2012ti}, where $L_{f\tilde g}\equiv L_\mu$. 
The NLO hard functions for gluino pair production in the momentum-space formalism then read
\begin{eqnarray}
H_{gg,\,{\bf I}}^{(1)}(\mu) &=&  
  4\,\*C_{1,\,{\bf I}}^{gg} - 12\,\*\CI 
    - C_A\*\left(64 - \frac{11\*\pi^2}{3}\right) \\
&&  - \Bigl(2\,\*\CI - 8\,\*C_A\*\ln(2)\Bigr)\*L_\mu 
    - 2\,\*C_A\*L_\mu^2 \non \\[2mm]
H_{\qqbar,\,{\bf 8}_a}^{(1)}(\mu) &=& 
  4\,\* C_{1,\,{\bf 8}_a}^{\qqbar} - 12\,\*C_A 
    - C_F\*\left(64 - \frac{11\*\pi^2}{3}\right) \\
&&  - \Bigl(2\,\*C_A - 8\,\*C_F\*\ln(2) -2\*\beta_0 + 6\,\*C_F\Bigr)\*L_\mu 
    - 2\,\*C_F\*L_\mu^2 \non
\end{eqnarray} 
in the $\overline{\text{MS}}$ scheme with $n_l=5$ active light (mass-less)
quark flavors. 
In the gluon fusion channel, only color-symmetric parts contribute near the threshold. 
Therefore, one has to sum over ${\bf I}\,={\bf 1},\,{\bf 8},\,{\bf 27}$. 
On the other hand, in the quark-antiquark annihilation
channel, one only has contributions from the anti-symmetric octet to first approximation.
The one-loop matching constants $C_{1,\,{\bf I}}^{gg}$ and $C_{1,\,{\bf 8}_a}^{\qqbar}$
are given in Eqs.~(39) and (40) of Ref.~\cite{Langenfeld:2012ti}.
They are functions of the ratio $r=\mq^2/\mg^2$ of the squared squark and
gluino masses.

If one expands the resummation formula (\ref{eq:sigmares2}) up to $\ord(\as^2)$, 
one obtains some constants besides the threshold logarithms at NNLO.
In order to obtain the full NNLO hard coefficients $H_{ij,\,{\bf I}}^{(2)}$, a dedicated 
two-loop calculation is necessary to determine all constant terms.
As these are not available at the moment, it is common to keep only threshold enhanced terms.
Thus, if one wishes to match the NNLL resummation onto the NNLO approximation, one requires 
a cancellation of all constants in the resummation formulas at $\ord(\as^2)$.  
Within the Mellin-space approach, this is achieved by a proper choice of 
$g^0_{ij,\, {\bf I}}$ in Eq.~(\ref{eq:sigmaNres}), see Ref.~\cite{Langenfeld:2012ti}.
For an analogous treatment in the momentum-space approach, we need to compute 
the next order in Eq.~(\ref{eq:sigmaresNLO}).
Keeping only the $\ord(a_s^0)$ contribution of the potential function, $\mg^2\beta/(2\pi)$,
and the hard function up to $\ord(a_s)$, the convolution with the NLO soft function and 
multiplication with the evolution function reproduces the soft logarithms $\ln^n(\beta)\ (n=1,..,4)$ 
and their coefficients at NNLO if, and only if, we choose $\mus=k_s\mg\beta^2$.
The explicit fixed-order expressions can be found in Ref.~\cite{Beneke:2009ye,Langenfeld:2012ti}. 
The NNLO coefficient of the hard function has to be chosen such
that the constants of the above expansion cancel at $\ord(\as^2)$. Here, one should keep in mind 
that also the yet unknown NNLO soft function, if taken into account, produces constant pieces at 
$\ord(a_s^2)$. In our case, we just arrange the NNLO hard function in such a way that all NNLO 
constant pieces cancel.
This ensures a proper matching on the approximated NNLO cross section
under usage of the NLO soft function. The explicit result is given in App.~\ref{sec:h2}. 
The relative impact on the total hard function is rather small, typically at the percent level. 
Finally, we mention that for NNLL+$\text{NNLO}_{\rm approx}$ accuracy, one can also set 
$H_{ij,\,{\bf I}}^{(2)}=0$, if the matching to the full fixed order 
(see discussion in the next section) is performed at the NNLO level. 
This differs from the choice in Eq.~(\ref{eq:H2match}) only
in higher order terms beyond NNLL accuracy, and is applied in Ref.~\cite{Beneke:2011mq}.
For NNLL resummation, all input functions to the resummation 
formula are needed at $\ord(\as)$. Note that the NLO expression of the
Sudakov exponent $S(\nu,\mu)$ requires the coefficients of the cusp anomalous 
dimension and the QCD beta function up to NNLO (see Ref.~\cite{Becher:2007ty}). 
All other anomalous dimension functions are required at NLO only.

\section{Implementation of the Mellin-space formalism}
\label{sec:implN}

In this section, we discuss the numerical implementation of the
resummation formula (\ref{eq:sigmaNres}).
The following procedure is required in general \cite{Bonciani:1998vc}:
First, as contributions of higher powers in $\beta$ become numerically
important when calculating the inclusive cross section, one has to match
the resummed cross section to the full fixed order result.
This is achieved by adding the latter, while subtracting all terms up to $\ord(a_s)$
(modulo the common prefactor $\alpha_s^2$) of the expanded NNLL expression. 
In order to perform the matching on the $\text{NNLO}_{\rm approx}$ partonic cross section, 
one needs to subtract all terms up to $\ord(a_s^2)$ of expanded resummation 
formula (denoted by NNLL(2)) and add back the approximated NNLO result \cite{Beneke:2011mq}.  

If one chooses to resum in Mellin space, an inverse Mellin transformation is
required which has to be done numerically. 
Due to omitted $1/N$ contributions in the resummation formula (\ref{eq:sigmaNres}), the
matching onto the NNLO approximation will give a slightly different result compared to
the NLO matching\footnote{Here, one has to emphasize that the numerically inverted NNLL(2) cross
section does not exactly correspond to threshold terms $\text{NNLO}_{\rm th}$ in momentum
space, but comes close towards the threshold.}.  
The resummed partonic cross section is given by
\begin{eqnarray}
\label{eq:sigmamatch}
\hat{\sigma}_{ij \to \gluinopair}^{{\rm res,NNLO}_{\rm approx}} 
&=& {\rm\bf M}^{-1}\Bigl[\hat{\sigma}_{ij \to \gluinopair}^{\rm NNLL,\,N} 
        - \hat{\sigma}_{ij \to \gluinopair}^{\rm NNLL(2),\,N}\Bigr]
    + \hat{\sigma}_{ij \to \gluinopair}^{{\rm NNLO}_{\rm approx}} \,.
\end{eqnarray}

\begin{figure}
\centering
 \scalebox{1}{\includegraphics{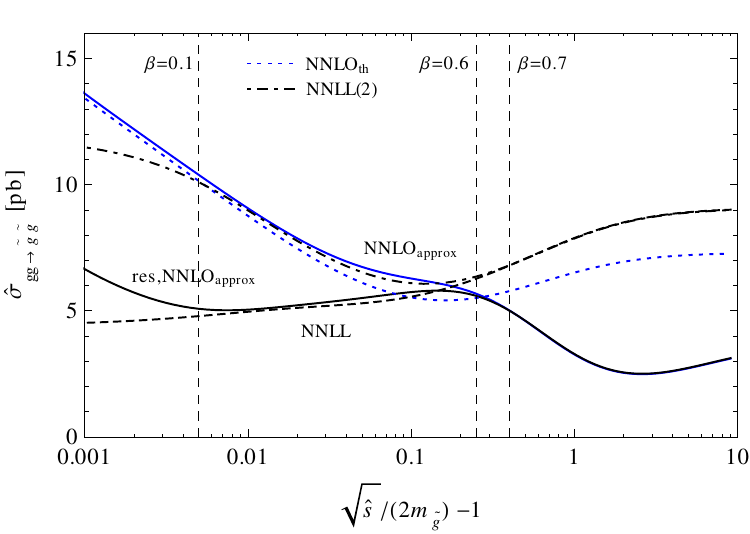}}\\
\vspace{3mm}
 \scalebox{1}{\includegraphics{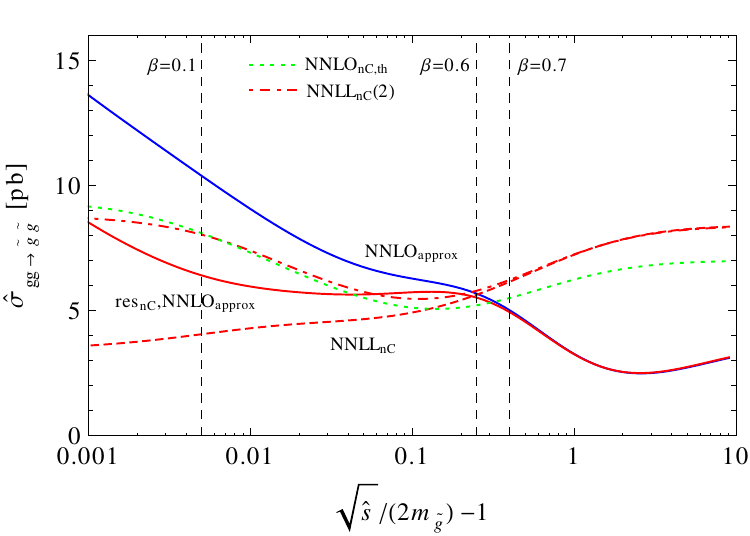}}\\
\vspace{3mm}
\caption{Partonic cross section for gluon fusion versus the energy above the
production threshold normalized to the gluino pair mass. The renormalization and
factorization scales have been set to $\mu=\mg=800\GeV$. 
We plot the NNLO approximation which is exact up to NLO,
and the NNLO threshold limit $\text{NNLO}_{\rm th}$ which contains only
threshold-enhanced contributions and NLO constants. Moreover, we show the NNLL
resummed cross section and its expansion in $\as$ up to second order, NNLL(2), 
as well as the resummed partonic cross section matched onto the
NNLO approximation. First panel: 
Coulomb corrections are included at fixed order in the resummation formula. 
Second panel: Coulomb corrections are neglected during resummation but kept in the 
NNLO approximation.}
\label{fig:sggpart}
\end{figure}

The numerical Mellin inversion requires an analytic continuation of the resummed cross
section to the complex plane. In our case, one simply assumes complex arguments of the logarithms.
A more general discussion on that topic is provided in 
\cite{Blumlein:1998if,Blumlein:2000hw}, and a good routine is included in the 
program \texttt{ANCONT} \cite{Blumlein:2000hw}.
Here, the Mellin inversion is obtained by
\begin{eqnarray}
f(\rho)=\frac 1\pi \int_0^\infty dz\,{\rm Im}\bigl[ e^{i\Phi}\rho^{-c(z)}
      {\rm\bf M}[f](c(z))\bigr]\,,
\end{eqnarray}  
where Mellin-$N$ is identified with $c(z)$, and $c(z)=c_0+z e^{i\Phi}$.
As the contour integral around the singularity at $N=0$ is symmetric with respect to 
the $x$-axis, only the upper half is evaluated and multiplied by two. 
The path is chosen to be a truncated line with an angle $\Phi$ close to $\pi$. 
The parameter integral over $z$ is divided logarithmically into 20 pieces, where 
each segment is performed by the 32-point Gauss formula.
We choose the starting point $c_0=1.4$ to the right of the $1/N$-pole,
but left to the Landau pole singularity which is excluded from the integration
contour according to the minimal prescription presented in Ref.~\cite{Catani:1996dj,Catani:1996yz}.

\begin{figure}
\centering
\scalebox{1}{\includegraphics{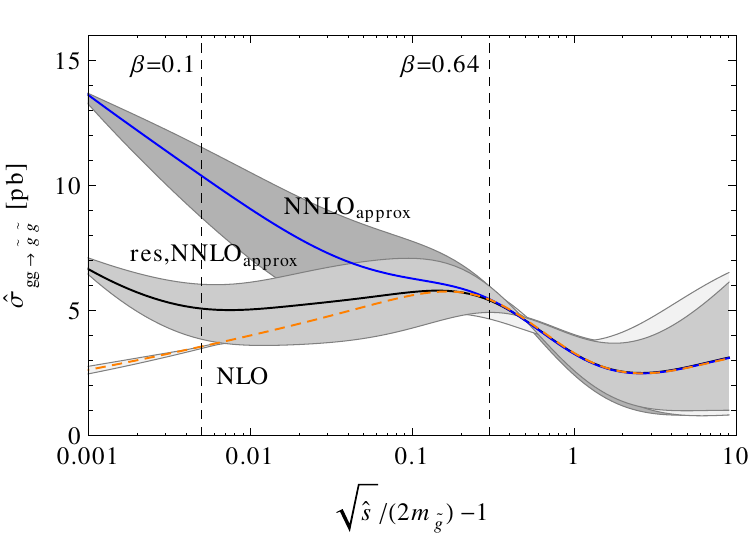}}
\caption{Comparison of the resummed partonic cross section to the exact NLO result
and the NNLO approximation. 
The scale is varied within the interval $\mu\in[\mg/2,2\mg]$ around $\mu=\mg=800\GeV$. }
\label{fig:sggpartcomp}
\end{figure}

In Fig.~\ref{fig:sggpart}, we plot the resummed partonic cross section in the gluon fusion 
channel against the energy above the threshold normalized to the
gluino pair mass $E/(2\mg)=\sqrt{\hat s}/(2\mg) -1$ (solid black). 
The Coulomb corrections are included at fixed NNLO, and we set $\mu=\mg$. 
We also plot the NNLO approximation (solid blue) and the threshold approximation
 $\text{NNLO}_{\rm th}$ (dotted blue).
Concerning the latter, all analytical input is given in Ref.~\cite{Langenfeld:2012ti}.
As a check of the numerical Mellin inversion, we further show the inverted expansion of the 
soft resummation NNLL(2) which comes close to $\text{NNLO}_{\rm th}$ and the NNLO approximation
for $\beta<0.6$. The inverted soft resummation NNLL merges with its $\ord(\as^2)$ 
expansion away from the threshold.
For low velocities ($\beta\ll0.1$), the numerical quality of the Mellin inversion decreases. 
As this systematic error affects both NNLL and NNLL(2), one may expect
a cancellation in the difference to some extent. One the other hand, $\beta< 0.1$ corresponds 
to an energy region $E< 10\GeV$ for $\mg\approx 1\TeV$
which is of minor importance for the evaluation of the total hadronic cross section.
In the second panel, the resummation is done by omitting all Coulomb corrections (these
are still included in the NNLO approximation).
In Fig.~\ref{fig:sggpartcomp}, we compare the resummed cross section to the exact NLO result,
where we also plot the theoretical error related to scale 
variation in the interval $\mu\in[\mg/2,2\mg]$.

As evident from the figures, the NNLL soft resummation significantly depletes the partonic cross 
section compared to the NNLO  approximation and brings it down near the
level of the NLO calculation. Close to the threshold, the resummed cross section is enhanced compared 
to fixed NLO but still well below the NNLO approximation.
In the kinematical region $\beta\approx 0.6$, the resummed cross section, the 
NNLO approxi\-mation, and the threshold limit $\text{NNLO}_{\rm th}$ cross each other close by.
For higher velocities ($\beta>0.7$), threshold approximations become untrustworthy,
but cancel within Eq.~(\ref{eq:sigmamatch}).
This gives rise to a smooth transition between the fixed-order and the resummed partonic cross section.
Note that for the gluon fusion channel, the results are rather stable against a variation
of the squark masses which enter at NLO. In constructing Fig.~\ref{fig:sggpart}, we have
assumed as example $\mg=800\GeV$ and a common squark mass of $640\GeV$. 
Choosing squark masses above $1\TeV$ will not change
the conclusions drawn above. This is not true for the quark-antiquark annihilation
channel. However, the hadronic production cross section is mainly driven by gluon fusion
(about $99\%$). Thus, we will not have a closer look on $\qqbar$-annihilation.

\section{Implementation of the momentum-space formalism}
\label{sec:implbeta}

Within the momentum-space formalism, no inversion is required and one 
can remove the operator ${\rm\bf M}^{-1}$ in Eq.~(\ref{eq:sigmamatch}).
However, the $\w$-integral over the potential 
function in Eq.~(\ref{eq:sigmares2}) requires an analytic continuation to negative 
values of $\eta$, where the integrand has to be understood in a distributional 
sense such that the integral is convergent.
For NNLL resummation of gluino pairs we need to integrate 
\begin{eqnarray}
\label{eq:common}
I(\w) & = & \frac 1\w {\left(\frac{\w}{2\mg}\right)}^{2\eta}\,
\frac{e^{-2\gE\, \eta}}{\Gamma(2\eta)}
\left(I^{(0)}(\w) + 16 \pi^2 a_s^2\, I^{(0)S}(\w) + I^{(1)}(\w) \right)\,,
\end{eqnarray}
where
\begin{eqnarray}
\label{eq:I0}
I^{(0)}(\w) &=& J_{\bf I}^{(0)}\bigl(E-\w/2\bigr)
\Bigg{(}1 \,+\, 4\,a_s\Bigg{[} \CI\left(1 + \hat\psi(2\eta)
+ \ln\Bigl(\frac{\mus}{2\mg}\Bigr)\right)\\
&& \,+\, (C_i + C_j)\left(\frac{\pi^2}{24} + \hat\psi(2\eta)^2 
- \psi'(2\eta) + 2\, \hat\psi(2\eta) \ln\Bigl(\frac{\mus}{2\mg}\Bigr)
+ \ln^2\Bigl(\frac{\mus}{2\mg}\Bigr) \right) \non \\
&& - \bigg{\lbrace}\CI + 2(C_i + C_j)\left(\hat\psi(2\eta)
+ \ln\Bigl(\frac{\mus}{2\mg}\Bigr)\right)\bigg{\rbrace} \ln\left(\frac{\w}{2\mg}\right)
+ (C_i + C_j) \ln^2\left(\frac{\w}{2\mg}\right) \Bigg{]} \Bigg{)}\,,\non\\
\label{eq:I0S}
I^{(0)S}(\w) &=& J_{\bf I}^{(0)}\bigl(E-\w/2\bigr)\,
\ln\left(\frac{E-\w/2}{\mg}\right)\,,\\
\label{eq:I1}
I^{(1)}(\w) &=& J_{\bf I}^{(1)}\bigl(E-\w/2\bigr)\,.
\end{eqnarray}
Here, we have defined $\hat\psi(x) = \gE + \psi(x)$, where $\psi$ denotes the digamma function.
Above the threshold, the integral runs from $0$ to $2E$.
As outlined in Ref.~\cite{Becher:2006mr,Beneke:2010da} for instance, analytic
continuation to negative values of $\eta$ can be achieved by a replacement of
the integration kernel by a so-called star distribution which contains  
subtraction terms that render the integral finite for $\eta>-1$. 
Explicitly, one applies
\begin{eqnarray}
\label{eq:cstar1}
&&\hspace{-12mm}\int_0^{2 E}d\w\,f\bigl(E-\frac{\w}{2}\bigr)
\left[\frac 1\w {\left(\frac{\w}{2\mg}\right)}^{2\eta}\right]_*\\
& = & \int_0^{2 E} \frac{d\w}{\w} \Bigl[ f(E-\frac{\w}{2})-f(E)
+\frac\w 2 f'(E)\Bigr] 
{\left(\frac{\w}{2\mg}\right)}^{2\eta}
 +\, \left[\frac{f(E)}{2\eta}-\frac{f'(E)E}{2\eta+1}\right]
 {\left(\frac{E}{\mg}\right)}^{2\eta}, \non\\[2mm]
\label{eq:cstar2}
&&\hspace{-12mm}\int_0^{2 E}d\w\,f\bigl(E-\frac{\w}{2}\bigr)\left[\frac{\ln\bigl(\frac{\w}{2\mg}\bigr)}{\w}
{\left(\frac{\w}{2\mg}\right)}^{2\eta}\right]_* \\
& = & \int_0^{2 E} \frac{d\w}{\w} \Bigl[ f(E-\frac{\w}{2})-f(E)+\frac\w 2 f'(E)\Bigr] 
\ln\Bigl(\frac{\w}{2\mg}\Bigr) {\left(\frac{\w}{2\mg}\right)}^{2\eta}\non\\
&&  +\, \Biggl[\frac{f(E)}{2\eta}\left(L_E-\frac{1}{2\eta}\right)
-\frac{f'(E)E}{2\eta+1}\left(L_E-\frac{1}{2\eta+1}\right)\Biggr]
 {\left(\frac{E}{\mg}\right)}^{2\eta}, \non\\[2mm]
\label{eq:cstar3}
&&\hspace{-12mm}\int_0^{2 E}d\w\,f\bigl(E-\frac{\w}{2}\bigr)\left[\frac{\ln^2\bigl(\frac{\w}{2\mg}\bigr)}{\w}
{\left(\frac{\w}{2\mg}\right)}^{2\eta}\right]_* \\
& = & \int_0^{2 E} \frac{d\w}{\w} \Bigl[ f(E-\frac{\w}{2})-f(E)+\frac\w 2 f'(E)\Bigr] 
\ln^2\Bigl(\frac{\w}{2\mg}\Bigr) {\left(\frac{\w}{2\mg}\right)}^{2\eta}\non\\
&& +\, \Biggl[\frac{f(E)}{2\eta}\left(L_E^2
-\frac{1}{\eta}L_E+\frac{1}{2\eta^2}\right)
-\frac{f'(E)E}{2\eta+1}\left(L_E^2
-\frac{2}{2\eta+1}L_E+\frac{2}{(2\eta+1)^2}\right)\Biggr]
{\left(\frac{E}{\mg}\right)}^{2\eta}, \non
\end{eqnarray}
with $L_E\equiv \ln(E/\mg)\,$ and $f(E-\w/2)$ being a smooth test function on the interval $\w\in[0,2E]$. 
For positive $\eta$, one can drop the star-brackets
and the above relations become simple identities.
For negative $\eta$, diverging boundary terms at $\w=0$ are removed by the star
prescription. In deriving the fixed order expansion (\ref{eq:sigmaresNLO}),
we have assumed positive $\eta$ in order to perform the convolution over the LO
potential function. We want to stress here that there are different methods
for the analytic continuation. An alternative treatment, based on integration by parts, 
has been proposed in Ref.~\cite{Beneke:2011mq}. However, due to the
identity theorem for holomorphic functions, the continuation to $\eta<0$ is unique
as the prescriptions agree for positive $\eta$ \footnote{Here, I want to thank Pietro Falgari
for clarifying discussions.}.

A different ansatz for analytic continuation is required for the non-Coulomb corrections. 
Here, one may apply 
\begin{eqnarray}
\label{eq:cstarS}
&&\hspace{-12mm}\int_0^{2 E}d\w\,J_{\bf I}^{(0)}\bigl(E-\frac \w 2\bigr)\,
\ln\left(\frac{E-\w/2}{\mg}\right)
\left[\frac 1\w {\left(\frac{\w}{2\mg}\right)}^{2\eta}\right]_*\\
& = & \int_0^{2 E} \frac{d\w}{\w} \Bigl[ J_{\bf I}^{(0)}(E-\frac{\w}{2})
-J_{\bf I}^{(0)}(E)+\frac\w 2 J_{\bf I}^{(0)\,\prime}(E)\Bigr] 
{\left(\frac{\w}{2\mg}\right)}^{2\eta}\ln\left(\frac{E-\w/2}{\mg}\right) \non\\
&& +\, \Biggl[\frac{J_{\bf I}^{(0)}(E)}{2\eta}
\left(L_E-\gE-\psi(1+2\eta)\right) 
-\frac{J_{\bf I}^{(0)\,\prime}(E)E}{2\eta+1}
\left(L_E-\gE-\psi(2+2\eta)\right)\Biggr]
 {\left(\frac{E}{\mg}\right)}^{2\eta},\non
\end{eqnarray}
where the terms in the last two lines have been evaluated
for positive $\eta$. See also Ref.~\cite{Beneke:2011mq} for a slightly different treatment.
The latter relations can also be applied to the logarithm in Eq.~(\ref{eq:J1fix})
if one includes the NLO corrections from the Coulomb potential at fixed order.

The poles at $\eta=-1/2$ and $\eta=-1$ in Eqs.~(\ref{eq:cstar1}-\ref{eq:cstarS}) 
are canceled by the overall factor $1/\Gamma(2\eta)$ in formula (\ref{eq:common}), 
see also the discussion in Ref.~\cite{Becher:2006mr,Beneke:2010da}. 
The pole at $\eta=0$ is only canceled if there is no logarithm $\ln(\w)$ contained in the integral.
Otherwise, the prefactor $1/\Gamma(2\eta)$ is not sufficient to give a final result
for $\eta\to 0$ as can be seen on the right-hand sides of Eqs.~(\ref{eq:cstar2})
and (\ref{eq:cstar3}).
However, this does not cause any trouble. The limit $\eta\to 0$ 
corresponds to $\mus \to\muf$ which, using a running soft 
scale as stated above, is achieved for large velocities ($\beta\approx 1$). 
However, the factorization ansatz in Eq.~(\ref{eq:sigmafac}) is not valid in that 
kinematical region\footnote{The same statement is also true for the corresponding ansatz 
of the Mellin-space approach in Eq.~(\ref{eq:sigmaNres}), of course.}.
For instance, it is easy to see that some NLO terms in the soft function 
dominate the LO term, when $\eta$ approaches zero.
Thus, one should not apply resummation for $\beta$ close to one.
This will exclude the problematic case $\eta\to 0$. 
 
Another complication arises due to the terms proportional to 
$f(E)\,L_E^k\, E^{2\eta}$ with $k=0,1,2\,$ in the formulas~(\ref{eq:cstar1}-\ref{eq:cstar3}),
 and $J_{\bf I}^{(0)}(E)\,L_E\,E^{2\eta}$ in Eq.~(\ref{eq:cstarS}), which in most cases 
diverge for $E\to 0$. 
In our case, only the repulsive potential $J_{\bf 27}^{(0)}$ and its products with
powers of $\ln(E)$ evaluate to zero at the threshold. For $\eta<-1/2$, the
convolution with the parton luminosity is ill-defined, see also the equivalent
discussion about top-quark pair production in Ref.~\cite{Beneke:2011mq}.
However, as outlined in the latter reference, the continuation to negative values 
of $\eta$ can be extended to the convolution over the hadronic cms energy
producing a finite hadronic cross section.

At this point, we want to note that something seems to be different compared
to the treatment in Mellin-space.
Suppose that we neglect all kind of Coulomb corrections.
It follows that the fixed order partonic cross section has to vanish near the threshold due to
the overall factor $\beta$ in the born cross section. 
In Mellin space, this behavior is not spoiled by soft resummation.
In the momentum-space approach in the absence of Coulomb corrections, 
the potential function reduces to the zeroth order in Eq.(\ref{eq:JI0})
which is proportional to $\sqrt{E/\mg}\approx\beta$. 
If this is multiplied by $E^{2\eta}$, one finds a singular behavior for $E\to 0$ 
and $\eta<-1/4$ at the partonic level. 
Note that such a phenomenon does not occur in the discussion about deep inelastic scattering 
in Ref.~\cite{Becher:2006mr}, simply because the momentum integral over $p^2$ runs from zero 
to $Q^2$ with $Q$ being the hard scale of the problem.
  
In any case one should state that resummation defines the partonic cross section as a distribution only 
irrespective if Coulomb terms are resummed or not\footnote{Here, I want to thank
Martin Beneke and Christian Schwinn for clarifying discussions.}.
Therefore, there is no meaningful way to compare with the purely soft resummation
in the Mellin-space formalism at the partonic level, where the threshold behavior
mimics the fixed order in the absence of Coulomb terms, but adhoc 
assumptions have been made in order to deal with the Landau singularity. 

We close this discussion by noting that threshold resummation in general suffers from 
ambiguities dealing with the question of how to treat the Landau pole on the
one hand, or such related to the choice of scales and the matching procedure on the other.
The explicit implementation of the momentum-space formalism is passed over to the experts
on that approach.

\begin{figure}[t]
\centering
 \scalebox{0.64}{\includegraphics{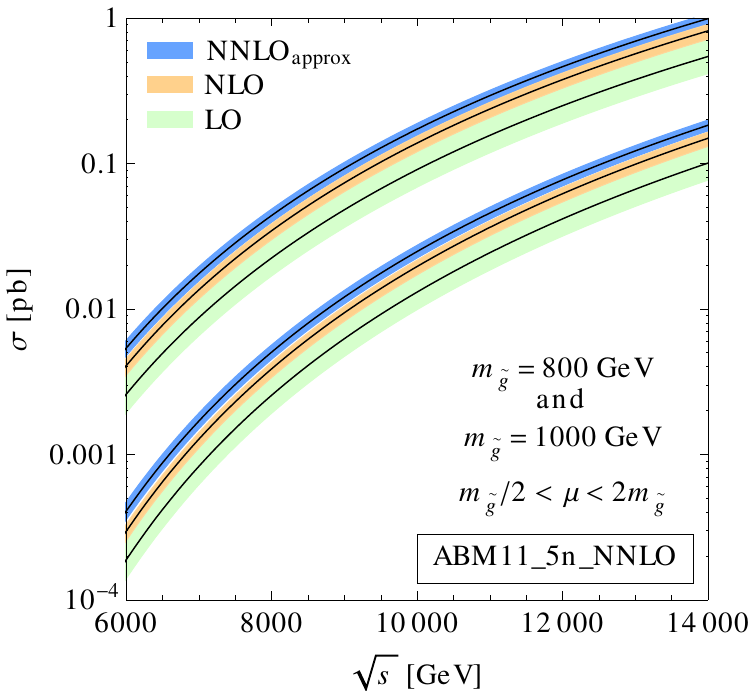}}
\hspace{1mm}
 \scalebox{0.64}{\includegraphics{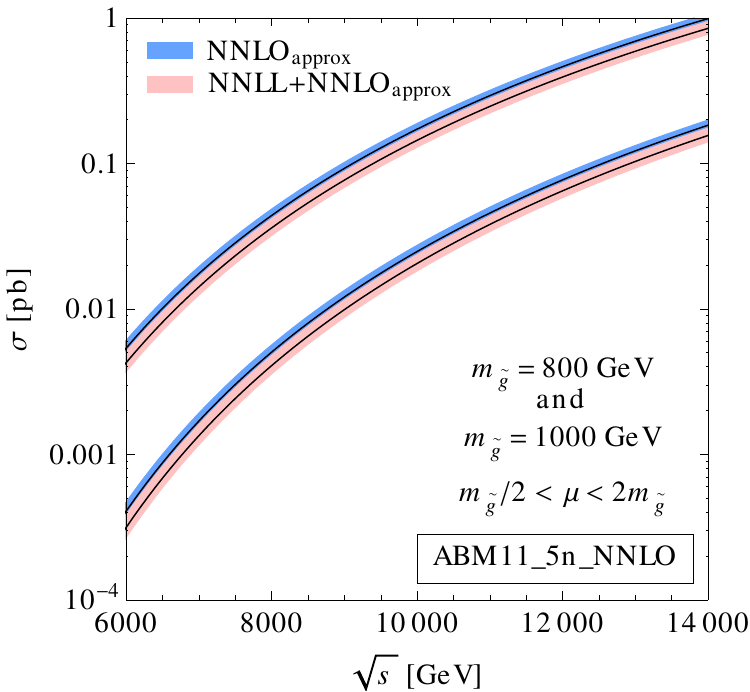}}
\caption{Inclusive hadronic cross section versus the squared cms energy. The error bars refer
only to scale variation. Left panel: Fixed order for $\mg=800\GeV$ and $\mg=1\TeV$. 
Right panel: NNLO approximation and soft NNLL resummation matched onto the latter.}
\label{fig:sigmahad}
\end{figure}

\section{Hadronic cross section}
\label{sec:pheno}

In this section, we examine the impact of the NNLL resummation onto the inclusive hadronic cross 
section for the examples $\mg=800\GeV$ and $\mg=1\TeV$.
The current exclusion bounds suggest a minimum gluino mass of about $1\TeV$ \cite{:2012rz,:2012gq}. 
However, we want to stress again that these results depend strongly on the chosen 
PDF set and the value of $\alpha_s(M_Z)$.
The libraries CTEQ6.6 \cite{Nadolsky:2008zw} and MSTW08 \cite{Martin:2009iq} used in 
Ref.~\cite{:2012rz} lead to similar results of the hadronic production cross section. 
The more recent sets ABM11 \cite{Alekhin:2012ig} for instance differ from the latter at 
large parton momentum fraction $x$ which is of special importance for heavy (s)particle 
production. As discussed in Ref.~\cite{Langenfeld:2012ti}, the usage of ABM11 with 
$\as(M_Z)=0.1134\pm0.0011$ compared to MSTW08 featuring $\as(M_Z)=0.1171\pm0.0014$
reduces the hadronic production cross section by a factor of about two.
Concerning the relative effect of the NNLL resummation compared to fixed NLO or
$\text{NNLO}_{\rm approx}$, the actual choice of the PDF set will only play a minor role. 
In our analysis, we compare the widely used MSTW08 with ABM11 PDFs, where we always use
NNLO sets also for the LO and NLO cross sections.
We further assume a fixed ratio $\mq/\mg=4/5$ giving rise to the values 
$\mq=640\GeV$ and $\mq=800\GeV$. 
As mentioned above, this choice has also no significant impact on the result. 
A variation of the squark masses would affect the inclusive hadronic cross section 
at the percent level only. 

\begin{table}[t]
\centering
\begin{tabular}{cccccc}
\toprule
$\mg\, [\GeV]$ & $\sqrt{s}\, [\TeV]$ & $\sigma^\text{LO} [\pb]$ & 
$\sigma^\text{NLO} [\pb]$ & $\sigma^{\text{NNLO}_{\rm approx}} [\pb]$ & 
$\sigma^{\text{res},\text{NNLO}_{\rm approx}} [\pb]$ \\
\midrule
\midrule
\multicolumn{6}{c}{MSTW 2008 NNLO}\\
\midrule
 & 7 & 0.0198 $\left({0.0293}\atop{0.0139}\right)$ & 
 0.0329 $\left({0.0385}\atop{0.0270}\right)$ & 
 0.0425 $\left({0.0477}\atop{0.0373}\right)$ & 
 0.0351 $\left({0.0407}\atop{0.0306}\right)$ \\
\cmidrule{2-6}  
800 & 8 & 0.0480 $\left({0.0697}\atop{0.0340}\right)$ & 
 0.0779 $\left({0.0900}\atop{0.0648}\right)$ &
 0.0989 $\left({0.1098}\atop{0.0877}\right)$ & 
 0.0828 $\left({0.0948}\atop{0.0729}\right)$ \\ 
\cmidrule{2-6} 
 & 14 & 0.9184 $\left({1.2699}\atop{0.6804}\right)$ &
 1.3953 $\left({1.5576}\atop{1.2041}\right)$ & 
 1.6853 $\left({1.8146}\atop{1.5406}\right)$ & 
 1.4636 $\left({1.6218}\atop{1.3321}\right)$\\
\midrule
 & 7 & 0.0020 $\left({0.0030}\atop{0.0014}\right)$ & 
 0.0034 $\left({0.0041}\atop{0.0028}\right)$ & 
 0.0046 $\left({0.0052}\atop{0.0039}\right)$ &
 0.0037 $\left({0.0044}\atop{0.0032}\right)$ \\ 
\cmidrule{2-6} 
1000 & 8 & 0.0059 $\left({0.0087}\atop{0.0041}\right)$ & 
 0.0097 $\left({0.0113}\atop{0.0080}\right)$ & 
 0.0126 $\left({0.0142}\atop{0.0110}\right)$ & 
 0.0104 $\left({0.0121}\atop{0.0090}\right)$ \\ 
\cmidrule{2-6} 
 & 14 & 0.1837 $\left({0.2564}\atop{0.1349}\right)$ & 
 0.2798 $\left({0.3131}\atop{0.2403}\right)$ & 
 0.3421 $\left({0.3709}\atop{0.3109}\right)$ & 
 0.2952 $\left({0.3294}\atop{0.2665}\right)$ \\
\midrule
\midrule
\multicolumn{6}{c}{ABM11 NNLO}\\
\midrule
 & 7 & 0.0087 $\left({0.0119}\atop{0.0065}\right)$ & 
 0.0135 $\left({0.0151}\atop{0.0117}\right)$ & 
 0.0175 $\left({0.0195}\atop{0.0156}\right)$ & 
 0.0142 $\left({0.0166}\atop{0.0125}\right)$ \\
\cmidrule{2-6} 
800 & 8 & 0.0224 $\left({0.0304}\atop{0.0168}\right)$ & 
 0.0345 $\left({0.0385}\atop{0.0299}\right)$ & 
 0.0439 $\left({0.0486}\atop{0.0394}\right)$ & 
 0.0361 $\left({0.0417}\atop{0.0320}\right)$ \\ 
\cmidrule{2-6} 
 & 14 & 0.5439 $\left({0.7289}\atop{0.4148}\right)$ & 
 0.8160 $\left({0.9031}\atop{0.7150}\right)$ & 
 0.9908 $\left({1.0674}\atop{0.9085}\right)$ & 
 0.8489 $\left({0.9459}\atop{0.7722}\right)$ \\ 
\midrule
 & 
 7 & 0.0008 $\left({0.0011}\atop{0.0006}\right)$ &
 0.0013 $\left({0.0014}\atop{0.0011}\right)$ & 
 0.0017 $\left({0.0019}\atop{0.0015}\right)$ & 
 0.0014 $\left({0.0016}\atop{0.0012}\right)$ \\ 
\cmidrule{2-6} 
1000 & 8 & 0.0025 $\left({0.0035}\atop{0.0019}\right)$ &
 0.0039 $\left({0.0043}\atop{0.0034}\right)$ & 
 0.0050 $\left({0.0057}\atop{0.0045}\right)$ &
 0.0041 $\left({0.0048}\atop{0.0036}\right)$ \\ 
\cmidrule{2-6} 
 & 14 & 0.1006 $\left({0.1346}\atop{0.0768}\right)$ & 
 0.1490 $\left({0.1641}\atop{0.1311}\right)$ & 
 0.1831 $\left({0.1986}\atop{0.1674}\right)$ & 
 0.1557 $\left({0.1750}\atop{0.1408}\right)$ \\
\bottomrule
\end{tabular}
    \caption{\small
      \label{tab:numbers}
      Hadronic cross section at $\sqrt{s}=7,8,$ and $14\TeV$, evaluated by using MSTW 2008 
      and ABM11 PDF sets . The ratio $\mq/\mg$ is kept fixed to the value $4/5$.
      The numbers in brackets correspond to the hadronic cross section evaluated at
      $\mu=\mg/2$ (upper number) and $\mu=2\mg$. PDF errors are not included. }
\end{table}

The theoretical error due to higher order corrections is estimated by a variation
of the scale $\mu=\mu_f=\mu_h$ in the range $\mu\in[\mg/2,2\mg]$.
Within the momentum-space approach, one further has the possibility for a variation
of the soft scale in order to account for the ambiguities within the soft resummation, 
as well as an independent variation of the hard scale.
In principle, the latter is also possible in the Mellin-space approach.
For illustration of the perturbative accuracy we will not include the PDF errors.
Plots showing these can be found in Ref.~\cite{Langenfeld:2012ti}.
The convolution of the (resummed) partonic cross sections with the parton luminosity
functions according to Eq.~(\ref{eq:totalcrs}) is performed by a \texttt{VEGAS} integration
routine using LHAPDF grid files \cite{Whalley:2005nh} for the PDF sets.

In the left panel of Fig.~\ref{fig:sigmahad}, we plot the fixed order inclusive cross section 
as a function of the hadronic cms energy $\sqrt{s}$ using ABM11 PDF sets.
Explicit numbers for $\sqrt{s}=7,8,$ and $14\TeV$ are given in Table~\ref{tab:numbers}.
For $\mg=800\GeV$ and $\sqrt{s}=7\TeV$, 
we observe the $K$-factors $K_{\rm NLO} = \sigma_{\rm NLO}/\sigma_{\rm LO} = 1.55$
and $K_{\rm NNLO} = \sigma_{\text{NNLO}_{\rm approx}}/\sigma_{\rm NLO} = 1.30$, 
at $14\TeV$ we have $K_{\rm NLO} = 1.50$ and $K_{\rm NNLO} = 1.21$.
For $\mg=1\TeV$ a similar picture emerges with $K_{\rm NLO} =1.63$ and $K_{\rm NNLO} =1.31$
for $\sqrt{s}=7\TeV$, and $K_{\rm NLO} =1.48$ and $K_{\rm NNLO} =1.23$ at $14\TeV$.

In the right panel of Fig.~\ref{fig:sigmahad}, we compare the NNLO approximation to the NNLL resummation, 
matched onto the latter. Explicit numbers are also given in Table~\ref{tab:numbers}.
In both examples, the $K$-factors $K_{\rm NNLL} = \sigma_{\text{res,NNLO}_{\rm approx}}/\sigma_{\rm NLO}$ 
are slightly above one ($K_{\rm NNLL} \approx 1.05$) over the whole range of hadronic cms energies. 
This is a direct consequence of the partonic result shown in Fig.~\ref{fig:sggpartcomp}, where after 
convolution with the parton luminosities the corrections due to resummation arise mainly from the 
region $\beta\in[0.1,0.6]$. 
Similar observations are made by usage of MSTW 2008 PDFs which can be seen by inspection
of Table~\ref{tab:numbers}. Here, one also notices that the central values exceed those
due to ABM11 by about a factor of two as mentioned above. As discussed in Ref.~\cite{Langenfeld:2012ti},
this discrepancy is neither covered by the individual PDF errors nor by the uncertainties due to 
scale variation which are given by the values in brackets. 

Concerning the impact of scale variation, we find that the uncertainty 
for the NNLL resummation resembles those of the NLO computation and the NNLO approximation
with a huge overlap to both cases.
This can also be understood from the partonic cross section shown in Fig.~\ref{fig:sggpartcomp},
where we compare the respective errors. Here one also notices that the partonic
fixed NLO result has a smaller error at low velocities not too close to the threshold, 
but also a smaller central value such that the relative contribution to the total
hadronic result is reduced in that kinematical region. 
We conclude that for the case of heavy gluino pairs, soft resummation suggests
that the fixed NLO description already gives a good theoretical estimate after convolution with the PDFs. 
An outstanding question is if this behavior is kept after Coulomb resummation which
is likely to be answered within the momentum-space approach soon.

\section{Conclusions}
\label{sec:conc}

In this article, we have discussed the effect of NNLL soft resummation on the 
inclusive cross section for gluino pair production at the LHC by the use
of the Mellin-space resummation formalism. Here, we have used analytic 
results of Ref.~\cite{Langenfeld:2012ti} and performed the Mellin inversion numerically 
by the use of standard methods.
 We have found that soft resummation compensates most of the enhancement of the NNLO soft 
logarithms compared to the fixed NLO result. This in turn suggests that
the NLO computation already provides a good theoretical input
for constructing exclusion limits at the LHC or, hopefully, to construct
confidence regions for future evidences.
At this point we want to stress again that the main source of uncertainty
is related to the non-perturbative input when different PDF sets and 
initial values for $\as(M_Z)$ are applied. As discussed in Ref.~\cite{Langenfeld:2012ti},
the individual PDF errors are not sufficient to cover the resulting discrepancies.

We further calculated the hard function required for
a joined soft and Coulomb resummation in the momentum-space approach.
Here, we recaptured the relevant points concerning the convolution over
the hadronic cms energy which are given in greater detail in the literature. 
Finally, we state that threshold resummation always suffers from ambiguities,
no matter which approach is applied.

\subsection*{Acknowledgments}

I would like to thank Ulrich Langenfeld and Sven-Olaf Moch for
good advice and also for providing me with useful pieces of code.
I further wish to thank Martin Beneke, Pietro Falgari, 
and Christian Schwinn for very useful discussions.
This work has been supported in part by the Deutsche Forschungsgemeinschaft in 
Sonderforschungs\-be\-reich/Transregio~9 and 
by the European Commission through contract PITN-GA-2010-264564 ({\it LHCPhenoNet}).

\begin{appendix}

\renewcommand{\theequation}{\ref{sec:h2}.\arabic{equation}}
\setcounter{equation}{0}
\section{Matching to the NNLO approximation}
\label{sec:h2}

In this appendix, we give the relevant part of the NNLO hard function which 
may be used for the matching onto the NNLO approximation. The specific choice given here 
sets all NNLO constants to zero within the resummation formula~(\ref{eq:sigmares2}). 
For the generic scale choice $\mus=k_s\mg\beta^2$, we obtain
\begin{eqnarray}
\label{eq:H2match}
&& H_{ij,\,{\bf I}}^{(2)match} = -C_{1,\,{\bf I}}^{ij} \Bigl(48 \CI\bigl(1 - \ln(2)\bigr) + 
     2 (C_i + C_j) \bigl(64 - 96 \ln(2) + 72 \ln^2(2) - \frac{11}{3} \pi^2\bigr)\Bigr) \\
&& + 144 \CI^2 \bigl(1 - \ln(2)\bigr) + 
  2 (C_i + C_j) \CI \bigl(384 - 480 \ln(2) + 216 \ln^2(2) + 
     11 \pi^2 \ln(2) - 22 \pi^2\bigr) \non\\
&& + 2 (C_i + C_j)^2 \bigl(512 - 768 \ln(2) + 576 \ln^2(2) + 44 \pi^2 \ln(2) - 33 \pi^2
  - \frac{176}{3} \pi^2 + \frac{121}{72} \pi^4\bigr) \non\\
&& + L_\mu \Bigl\lbrace -C_{1,\,{\bf I}}^{ij} \Bigl(8 \CI + 16 (C_i + C_j) \bigl(2 - 3 \ln(2)\bigr)\Bigr) 
   - 4 \CI \bigl(33 - 2 n_l - 3 \kappa_{ij}\bigr) \bigl(1 - \ln(2)\bigr) \non\\
&& + 24 \CI^2 - (C_i + C_j) \Bigl(\frac{1592}{3} - \frac 13 \CI \bigl(13 \pi^2 
   + 144 \ln(2) \bigl(1 - 2 \ln(2)\bigr)\bigr) + \frac{n_l}{9} \bigl(-272 + 408 \ln(2) \non\\
&& - 216 \ln^2(2) + 11 \pi^2\bigr) + \kappa_{ij}\bigl(-32 + 48 \ln(2) - 36 \ln^2(2) 
   + \frac{11}{6} \pi^2\bigr) - 268 \ln(2) - 528 \ln(2) \non\\
&& + 396 \ln^2(2) + 12 \pi^2 \ln(2) - \frac{169}{6} \pi^2\Bigr) 
   - \frac{2}{3} (C_i + C_j)^2 \bigl(768 - 960 \ln(2) + 1008 \ln^2(2) \non\\
&& - 432 \ln(2)^3 + 61 \pi^2 \ln(2) - 48 \pi^2 - 336 \zeta(3)\bigr) 
   + \gamma_i^{\phi(1)} + \gamma_j^{\phi(1)} + \gamma_{ij,\,{\bf I}}^{V(1)} \Bigr\rbrace 
   + L_\mu^2 \Bigl\lbrace-4 C_{1,\,{\bf I}}^{ij} (C_i + C_j)\non\\
&& - \CI \bigl(11 - \frac 23 n_l - 2 \kappa_{ij}\bigr) 
   + 2 \CI^2 + (C_i + C_j) \Bigl(-\frac{199}{3} 
   + 4 \CI \bigl(3 - 2 \ln(2)\bigr) + \frac 29 n_l \bigl(17 - 18 \ln(2)\bigr) \non\\
&& + 4 \kappa_{ij} \bigl(2 - 3 \ln(2)\bigr) + 66 \ln(2) + \pi^2\Bigr) 
   + (C_i + C_j)^2 \bigl(-32 + 64 \ln(2) - 24 \ln^2(2) + \frac{13}{6} \pi^2\bigr)\Bigr\rbrace \non\\
&& + L_\mu^3 \Bigl\lbrace(C_i + C_j) (-\frac{11}{3} + 2 \CI + \frac 29 n_l + \kappa_{ij}) - 
     4 (C_i + C_j)^2 \ln(2)\Bigr\rbrace + \frac 12 (C_i + C_j)^2 L_\mu^4\,, \non\\
&& + \ln(k_s) \Bigl\lbrace-(\gamma_i^{\phi(1)} + \gamma_j^{\phi(1)}) - \gamma_{ij,\,{\bf I}}^{V(1)}
   + 4 \CI \bigl(33 - 2 n_l\bigr) \bigl(1 - \ln(2)\bigr) + 24 \CI^2 \bigl(1 - \ln(2)\bigr) \non\\
&& + \frac 12 (C_i + C_j) \Bigl(\frac{3184}{3} + 
    \frac 23 \CI \bigl(672 - 1008 \ln(2) + 648 \ln(2)^2 - 35 \pi^2 \bigr) - 
    \frac 29 n_l \bigl(272 - 408 \ln(2)  \non\\
&& + 216 \ln(2)^2 - 11 \pi^2\bigr) - 1592 \ln(2) + 792 \ln(2)^2 + 24 \pi^2 \ln(2) 
   - \frac{169}{3} \pi^2 \Bigr) \non\\
&& + \frac 23 (C_i + C_j)^2 \Bigl(1152 - 1728 \ln(2) + 1296 \ln(2)^2 - 
    648 \ln(2)^3 + 105 \pi^2 \ln(2) - 70 \pi^2 - 336 \zeta_3 \Bigr)\Bigr\rbrace \non\\
&& + \ln^2(k_s) \Bigl\lbrace 2 \CI^2 + \CI (11 - \frac 23 n_l) + 
   \frac 12 (C_i + C_j) \Bigl(\frac{398}{3} + 8 \CI \bigl(7 - 9 \ln(2)\bigr) 
   - n_l \Bigl(\frac{68}{9} - 8 \ln(2)\Bigr)\non\\
&& - 132 \ln(2) - 2 \pi^2 \Bigr) 
   + \frac 16 (C_i + C_j)^2 \bigl(576 - 864 \ln(2) + 648 \ln(2)^2 - 35 \pi^2 \bigr)\Bigr\rbrace \non\\
&& + \ln^3(k_s) \Bigl\lbrace \frac 12 (C_i + C_j) \Bigl(\frac{22}{3} + 4 \CI - \frac{4}{9} n_l\Bigr) + 
  4 (C_i + C_j)^2 \bigl(2 - 3 \ln(2)\bigr) \Bigr\rbrace + \ln^4(k_s) \frac 12(C_i + C_j)^2\,, \non
\end{eqnarray}
where $\kappa_{gg}=0$ and $\kappa_{\qqbar}=-2\beta_0 + 6 C_F$. The relative correction to the NLO hard 
function is about a few percent only. 

\end{appendix}

\begin{footnotesize}


\begin{thebibliography}{10}


\bibitem{:2012rz}
G.~Aad {\it et al.}  [ATLAS Collaboration],
\newblock Phys.\ Rev.\ D {\bf 87} (2013), arXiv:1208.0949 [hep-ex].

\bibitem{:2012gq}
S.~Chatrchyan {\it et al.}  [CMS Collaboration], arXiv:1212.6961 [hep-ex].

\bibitem{Beenakker:1996ch}
W.~Beenakker, R.~H{\"o}pker, M.~Spira, and P.~M. Zerwas,
\newblock Nucl. Phys. {\bf B492}, 51 (1997), arXiv:hep-ph/9610490.

\bibitem{Beenakker:1996ed}
W.~Beenakker, R.~H{\"o}pker, and M.~Spira,
\newblock (1996), arXiv:hep-ph/9611232.

\bibitem{Sterman:1986aj}
G.~F. Sterman,
\newblock Nucl.Phys. {\bf B281}, 310 (1987).

\bibitem{Catani:1989ne}
S.~Catani and L.~Trentadue,
\newblock Nucl.Phys. {\bf B327}, 323 (1989).

\bibitem{Becher:2006nr}
T.~Becher and M.~Neubert,
\newblock Phys.\ Rev.\ Lett.\  {\bf 97} (2006) 082001, arXiv:hep-ph/0605050.

\bibitem{Becher:2006mr}
T.~Becher, M.~Neubert and B.~D.~Pecjak,
\newblock JHEP {\bf 0701} (2007) 076, arXiv:hep-ph/0607228].

\bibitem{Beneke:2010da}
M.~Beneke, P.~Falgari, and C.~Schwinn,
\newblock Nucl.Phys. {\bf B842}, 414 (2011), arXiv:1007.5414.

\bibitem{Falgari:2012hx}
P.~Falgari, C.~Schwinn, and C.~Wever,
\newblock JHEP {\bf 1206}, 052 (2012), arXiv:1202.2260.

\bibitem{Kulesza:2009kq}
A.~Kulesza and L.~Motyka,
\newblock Phys. Rev. {\bf D80}, 095004 (2009), arXiv:arXiv:0905.4749.

\bibitem{Beenakker:2009ha}
W.~Beenakker {\em et~al.},
\newblock JHEP {\bf 0912}, 041 (2009), arXiv:0909.4418.

\bibitem{Langenfeld:2012ti}
U.~Langenfeld, S.~-O.~Moch and T.~Pfoh,
\newblock JHEP {\bf 1211} (2012) 070, arXiv:1208.4281.

\bibitem{Hagiwara:2009hq}
K.~Hagiwara and H.~Yokoya,
\newblock JHEP {\bf 0910} (2009) 049, arXiv:0909.3204.

\bibitem{Kauth:2011vg}
M.~R. Kauth, J.~H. K{\"u}hn, P.~Marquard, and M.~Steinhauser,
\newblock Nucl.\ Phys.\ B {\bf 857} (2012) 28, arXiv:1108.0361.

\bibitem{Falgari:2012sq}
P.~Falgari, C.~Schwinn and C.~Wever,
\newblock (2012), arXiv:1211.3408.

\bibitem{Catani:1990rp}
S.~Catani and L.~Trentadue,
\newblock Nucl.Phys. {\bf B353}, 183 (1991).

\bibitem{Contopanagos:1996nh}
H.~Contopanagos, E.~Laenen, and G.~F. Sterman,
\newblock Nucl. Phys. {\bf B484}, 303 (1997), arXiv:hep-ph/9604313.

\bibitem{Catani:1996dj}
  S.~Catani, M.~L.~Mangano, P.~Nason and L.~Trentadue,
  Phys.\ Lett.\ B {\bf 378} (1996) 329
  [hep-ph/9602208].

\bibitem{Catani:1996yz}
S.~Catani, M.~L. Mangano, P.~Nason, and L.~Trentadue,
\newblock Nucl. Phys. {\bf B478}, 273 (1996), arXiv:hep-ph/9604351.

\bibitem{Kidonakis:1997gm}
N.~Kidonakis and G.~F. Sterman,
\newblock Nucl. Phys. {\bf B505}, 321 (1997), arXiv:hep-ph/9705234.

\bibitem{Moch:2005ba}
S.~Moch, J.~Vermaseren, and A.~Vogt,
\newblock Nucl.Phys. {\bf B726}, 317 (2005), arXiv:hep-ph/0506288.

\bibitem{Kawamura:2012cr}
H.~Kawamura, N.~Lo~Presti, S.~Moch, and A.~Vogt,
\newblock Nucl.\ Phys.\ B {\bf 864} (2012) 399, arXiv:1205.5727.

\bibitem{Beneke:2009ye}
M.~Beneke {\em et~al.},
\newblock Phys.Lett. {\bf B690}, 483 (2010), arXiv:arXiv:0911.5166.

\bibitem{Czarnecki:1997vz}
A.~Czarnecki and K.~Melnikov,
\newblock Phys.Rev.Lett. {\bf 80}, 2531 (1998), arXiv:hep-ph/9712222.

\bibitem{Beneke:1999qg}
M.~Beneke, A.~Signer, and V.~A. Smirnov,
\newblock Phys.Lett. {\bf B454}, 137 (1999), arXiv:hep-ph/9903260.

\bibitem{Czarnecki:2001gi}
A.~Czarnecki and K.~Melnikov,
\newblock Phys.Rev. {\bf D65}, 051501 (2002), arXiv:hep-ph/0108233.

\bibitem{Pineda:2006ri}
A.~Pineda and A.~Signer,
\newblock Nucl.Phys. {\bf B762}, 67 (2007), arXiv:hep-ph/0607239.

\bibitem{Beneke:2011mq}
M.~Beneke, P.~Falgari, S.~Klein, and C.~Schwinn,
\newblock Nucl.Phys. {\bf B855}, 695 (2012), arXiv:1109.1536.

\bibitem{Becher:2007ty}
T.~Becher, M.~Neubert and G.~Xu,
\newblock JHEP {\bf 0807} (2008) 030, arXiv:0710.0680.

\bibitem{Moch:2008qy}
S.~Moch and P.~Uwer,
\newblock Phys. Rev. {\bf D78}, 034003 (2008), arXiv:0804.1476.

\bibitem{Becher:2009qa}
T.~Becher and M.~Neubert,
\newblock JHEP {\bf 0906} (2009) 081, arXiv:0903.1126.

\bibitem{Beneke:2009rj}
M.~Beneke, P.~Falgari, and C.~Schwinn,
\newblock Nucl.Phys. {\bf B828}, 69 (2010), arXiv:0907.1443.

\bibitem{Bonvini:2013td}
M.~Bonvini, S.~Forte, M.~Ghezzi and G.~Ridolfi,
\newblock arXiv:1301.4502.

\bibitem{Fadin:1990wx}
V.~S. Fadin, V.~A. Khoze, and T.~Sjostrand,
\newblock Z.Phys. {\bf C48}, 613 (1990).

\bibitem{Beneke:2005hg}
M.~Beneke, Y.~Kiyo and K.~Schuller,
\newblock Nucl.\ Phys.\ B {\bf 714} (2005) 67, arXiv:hep-ph/0501289.

\bibitem{Bonciani:1998vc}
R.~Bonciani {\it et al.}
\newblock Nucl.\ Phys.\ B {\bf 529} (1998) 424
[Erratum-ibid.\ B {\bf 803} (2008) 234], arXiv:hep-ph/9801375.

\bibitem{Blumlein:2000hw}
J.~Blumlein,
\newblock Comput.\ Phys.\ Commun.\  {\bf 133} (2000) 76.

\bibitem{Blumlein:1998if}
J.~Blumlein and S.~Kurth,
\newblock Phys.\ Rev.\ D {\bf 60} (1999) 014018.

\bibitem{Nadolsky:2008zw}
P.~M.~Nadolsky {\it et al.}
\newblock Phys.\ Rev.\ D {\bf 78} (2008) 013004, arXiv:0802.0007.

\bibitem{Martin:2009iq}
A.~Martin, W.~Stirling, R.~Thorne, and G.~Watt,
\newblock Eur.Phys.J. {\bf C63}, 189 (2009), arXiv:0901.0002.

\bibitem{Alekhin:2012ig}
S.~Alekhin, J.~Bl{\"u}mlein, and S.~Moch,
\newblock Phys.\ Rev.\ D {\bf 86} (2012) 054009, arXiv:1202.2281.

\bibitem{Whalley:2005nh}
M.~R.~Whalley, D.~Bourilkov and R.~C.~Group,
\newblock arXiv:hep-ph/0508110.


\end{thebibliography}

\end{footnotesize}

\end{document}